\documentclass[twocolumn]{emulateapj}
\usepackage{amsbsy}

\newcommand{\hour}{\mbox{$^{\rm h}$}}

\newcommand{\minute}{\mbox{$^{\rm m}$}}
\newcommand{\second}{\mbox{$^{\rm s}$}}

\slugcomment{submitted to ApJ Letters}
\shorttitle{Submillimeter Polarization Spectrum of OMC-1}
\shortauthors{Vaillancourt et al.}

\begin{document}

\title{New Results on the Submillimeter Polarization Spectrum\\of the
  Orion Molecular Cloud}

\author{
John~E. Vaillancourt,\altaffilmark{1}
C.~Darren Dowell,\altaffilmark{1,2}
Roger~H. Hildebrand,\altaffilmark{3,4}
Larry~Kirby,\altaffilmark{3}
Megan~M. Krejny,\altaffilmark{5}
Hua-bai~Li,\altaffilmark{6}
Giles~Novak,\altaffilmark{5}
Martin~Houde,\altaffilmark{7}
Hiroko~Shinnaga,\altaffilmark{8}
and
Michael~Attard\altaffilmark{7}
}

\altaffiltext{1}{Division of Physics, Mathematics, \& Astronomy,
  California Institute of Technology, MS 320-47, 1200 E. California
  Blvd., Pasadena, CA 91125 \email{johnv@submm.caltech.edu}}
\altaffiltext{2}{also Jet Propulsion Laboratory}
\altaffiltext{3}{Enrico Fermi Institute and Department of Astronomy \&
  Astrophysics, University of Chicago, 5640 S. Ellis Ave., Chicago, IL
  60637}
\altaffiltext{4}{also Department of Physics}
\altaffiltext{5}{Department of Physics and Astronomy, Northwestern
  University, 2145 Sheridan Rd., Evanston, IL 60208}
\altaffiltext{6}{Harvard-Smithsonian Center for Astrophysics, 60 Garden St., MS-78, Cambridge, MA 02138}
\altaffiltext{7}{Department of Physics and Astronomy, University of
  Western Ontario, London, ON, Canada N6A 3K7}
\altaffiltext{8}{Caltech Submillimeter Observatory, 111 Nowelo St., Hilo, HI 96720}

\begin{abstract}
  We have used the SHARP polarimeter at the Caltech Submillimeter
  Observatory to map the polarization at wavelengths of 350 and 450
  $\micron$ in a $\sim 2\arcmin \times 3\arcmin$ region of the Orion
  Molecular Cloud.  The map covers the brightest region of the OMC-1
  ridge including the Kleinmann-Low (KL) nebula and the submillimeter
  source Orion-south.  The ratio of 450-to-350 $\micron$ polarization
  is $\sim 1.3\pm0.3$ in the outer parts of the cloud and drops by a
  factor of 2 towards KL\@. The outer cloud ratio is consistent with
  measurements in other clouds at similar wavelengths and 
  confirms previous measurements placing the minimum of the
  polarization ratio in dusty molecular clouds at $\lambda \sim
  350\,\micron$.
\end{abstract}

\keywords{
dust, extinction ---
ISM: clouds ---
ISM: individual (OMC-1) --- 
polarization ---
submillimeter}


\section{Introduction}

Studies of the wavelength dependence of interstellar polarization are
common place in the near-visible region of the spectrum (e.g.,
\citealt{serkowski58,martin90}; \citealt*{martin99};
\citealt{whittet04}).  Such studies have put constraints on the
mechanisms of magnetic grain alignment
\citep{roberge04,alexreview1,alexreview2} as well as the properties of
dust grains responsible for interstellar extinction (e.g.,
\citealt{shape,andersson07,whittet08}; and references therein). There
are fewer studies of the wavelength dependence at far-infrared and
submillimeter wavelengths ($\sim 0.1$ -- 1 mm) where dust grains are
responsible for most of the observed emission in Galactic clouds.

The first studies of multi-wavelength far-infrared/\-sub\-milli\-meter
polarimetry found unexpected results. Rather than a featureless
spectrum due to a single population of dust grains, the polarization
spectrum is observed to fall from 60 to 100 to 350 $\micron$ before
rising again to 850 and 1300 $\micron$ \citep{pspec,mythesis,paris}.
This spectral structure is attributed to the existence of multiple
dust grain populations whose polarizability or alignment efficiency is
correlated with either the grain temperature, the spectral dependence
of the emissivity (the ``spectral index''), or a combination of
both.  However, the existing data are quite sparse in terms of both
wavelength coverage and the types of objects observed (bright Galactic
clouds). Therefore, rather than pursue observations with the ability
to test specific physical models, our immediate goal is to improve the
empirical description of the spectrum by measuring the polarization at
additional wavelengths between 100 and 850 $\micron$ which bracket the
observed minimum at $350\,\micron$.

To further this goal we have begun a campaign to measure the
polarization of Galactic clouds at both 350 and 450 $\micron$ using
SHARP, the SHARC-II polarimeter at the Caltech Submillimeter
Observatory on the summit of Mauna Kea.  While this is a fairly short
wavelength baseline, such measurements will further constrain the
location of the polarization minimum.  For example, one can simply ask
whether the minimum is less than, greater than, or approximately equal
to $350\,\micron$.  This will place additional constraints on the
possible range of polarizations, temperatures, and emissivities of the
constituent grain populations (e.g., \citealt{aod}).

We have carried out the first set of observations for this project
towards the Orion Molecular Cloud (OMC-1), a bright and well-studied
region of massive star formation (e.g., \citealt{houde04,
  johnstone99,lis98}).  In \S\ref{sec-obs} below we review the SHARP
instrument and the polarimetric/photometric observations.  Maps
are presented in \S\ref{sec-results} followed by a discussion of the
polarization spectrum in \S\ref{sec-disc}.

\section{Observations and Data Reduction} \label{sec-obs}

SHARP \citep{sharpspie2,sharpao,sharpspie1} is a fore-optics module
that installs onto the SHARC-II camera. Incident radiation is split
into two orthogonally polarized beams which are then imaged onto
opposite ends of the $12 \times 32$ pixel SHARC-II bolometer array.
The result is a dual-polarization $12 \times 12$ pixel polarimeter
with a $55\arcsec \times 55\arcsec$ field of view (FOV\@).  The
polarization is modulated by stepping a half-wave plate (HWP) at the
relative angles 0, 22.5, 45, and 67.5 degrees.

Within each FOV and HWP position, standard photometric beam-switching
was performed with a chop throw of $5\arcmin$ and a chop position
angle in the range $80\arcdeg$ -- $150\arcdeg$ east of north.  This is
repeated at positions separated by $50\arcsec$ in right ascension and
declination to build maps larger than the FOV\@.  For this work we
have generated maps of size $2\arcmin \times 3\arcmin$ encompassing
the 2 brightest submillimeter cores in OMC-1 (the Kleinmann-Low
nebula, hereafter KL; and the submillimeter source of \citet*{khw}
sometimes called Orion-south, hereafter KHW\@).  Observations at 350
and 450 $\micron$ were carried out on 2007 February 16 and 2006 December
5, respectively.

The generation of Stokes parameters, polarization amplitudes, and
position angles within each single set of 4 HWP angles follows the
same general procedures outlined by \citet{stokes} and \citet{primer}.
Maps of the linear Stokes parameters I, Q, and U are generated from
the dithered and stepped array positions by interpolating the data
onto a finer grid and coadding (e.g., \citealt{martinjohn}).

The polarization data presented in \S\ref{sec-results} have been
corrected for positive bias \citep{simmons85,plimits} and for measured
polarization efficiencies of 93\% and 98\% at 350 and 450 $\micron$,
respectively.  The data have also been corrected for instrument
polarizations of $\approx$ 0.3--0.5\% at $350\,\micron$ and
$\lessapprox$ 0.2\% at $450\,\micron$ (\citealt{sharpao}; Vaillancourt
et al., in preparation).  The polarization position angle calibration,
measured by illuminating SHARP with an unpolarized source through a
calibration grid, is accurate to within $2\arcdeg$.

We obtained photometric data at $350\,\micron$ using SHARC-II in
camera mode on 2007 August 11--12.  Four raster scans (without
chopping) centered on Ori IRC2 were performed to cover a $\sim
10\arcmin \times 10\arcmin$ field.  Flux maps were produced using the
{\tt
  sharcsolve}\footnote{\url{http://www.submm.caltech.edu/$\sim$sharc/analysis/overview.htm}.
  No significant differences were seen between maps reduced using {\tt
    sharcsolve} and those reduced using {\tt CRUSH}
  (\url{http://www.submm.caltech.edu/$\sim$sharc/crush/}).} utility,
which models the source flux, sky background fluctuations, instrument
gains, and drifts in the instrument electronics.  Absolute fluxes are
calibrated with respect to the standard sources L\,1551 and Mars
(using peak flux estimates of 45.2 and $3.80\times10^3$ Jy per
$9\arcsec$ FWHM beam,
respectively)\footnote{\url{http://www.submm.caltech.edu/$\sim$sharc/analysis/calibration.htm}}
and have an uncertainty of $\sim 20$\%.

\section{Results} \label{sec-results}

Figure \ref{fig-map1} shows the polarimetric and photometric maps in
the $\sim 2\arcmin \times 3\arcmin$ region covered by the polarization
observations. The size of the gaussian smoothing kernel used to
interpolate the $350\,\micron$ polarization data has been chosen so
that the resulting resolution matches that of the $450\,\micron$
polarization data ($13\arcsec$ FWHM\@).  Figures
\ref{fig-map1}\emph{a}--\emph{c} plot the 350 and 450~$\micron$
polarization results over a contour map of total flux measured at
$350\,\micron$.  At each wavelength we plot vectors at an interval of
$9\farcs5$ (Figs.\ \ref{fig-map1}\emph{a}--\emph{b}).

Figure \ref{fig-map1}\emph{a} shows the polarization results
with the length of the vector proportional to the measured
polarization amplitude and the position angle parallel to the
polarization vector. The most obvious feature of this map is the
decrease in the polarization amplitude (at both wavelengths) towards
the KL intensity peak, and to a lesser extent towards KHW\@. This so-called
``polarization hole'' effect in which the polarization drops towards
intensity peaks has been observed in these and other sources by
experiments at a range of far-infra\-red and sub\-milli\-meter
wavelengths
(e.g., \citealt{dasth,coppin2000,mwf01,archive,hertzarchive}).

In Figure \ref{fig-map1}\emph{b} we have rotated the polarization
vectors by $90\arcdeg$ to show the inferred magnetic field direction.
The plotted $\boldsymbol{B}$-vectors are drawn with a constant length
(not proportional to polarization amplitude) in order to more clearly
observe the position angle differences across the map and between the
two wavelengths.  The position angles exhibit a clockwise rotation
with increasing wavelength, differing by more than $25\arcdeg$ around
KL and the region immediately east (less than 1\% of the data exhibit
larger changes).  This angular rotation is much smaller to the west
and south of KL\@.  The larger variation of the position angle with
wavelength towards KL than along the ridge between KL and KHW is also
clearly seen when the data are compared with observations at
$850\,\micron$ (not shown, but see e.g., \citealt{brenda-pspec}).

The color scale of Figure \ref{fig-map1}\emph{c} shows the ratio of
the polarization amplitude at the two wavelengths, $P(450)/P(350)$.
Most points in the map exhibit a two-point polarization spectrum which
increases with wavelength, $P(450)/P(350) > 1$ (Fig.\ \ref{fig-hist}).
The most obvious exceptions are towards KL and the region to
its northwest.  KHW does not exhibit a similar drop in the
polarization ratio.

\setcounter{figure}{1}
\begin{figure}[t]
\plotone{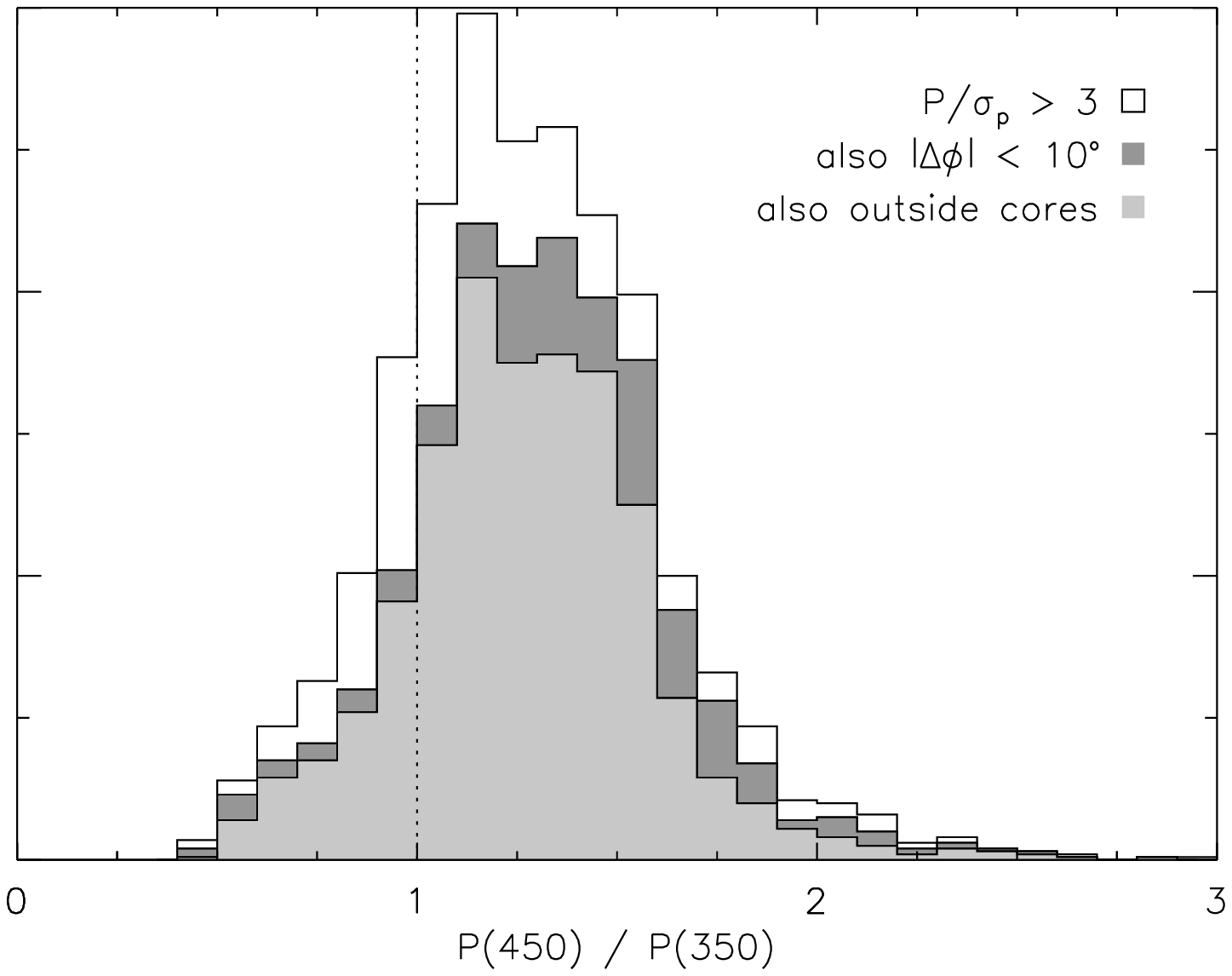}
\caption{Histogram of the 450/350 $\micron$ polarization ratio. All
  data shown here have been limited to only those points where $P \geq
  3\sigma_p$ at both wavelengths. Also shown are histograms where data
  points satisfy the additional criteria that the position angle
  rotates by less than $10\arcdeg$ between the two wavelengths
  ($|\Delta\phi| < 10\arcdeg$) and that the points are at least
  $20\arcsec$ away from the two submillimeter flux peaks KL and KHW.
}
\label{fig-hist}
\end{figure}

\section{Discussion} \label{sec-disc}

\subsection{Minimum in the Polarization Spectrum}

\begin{figure}[t]
\plotone{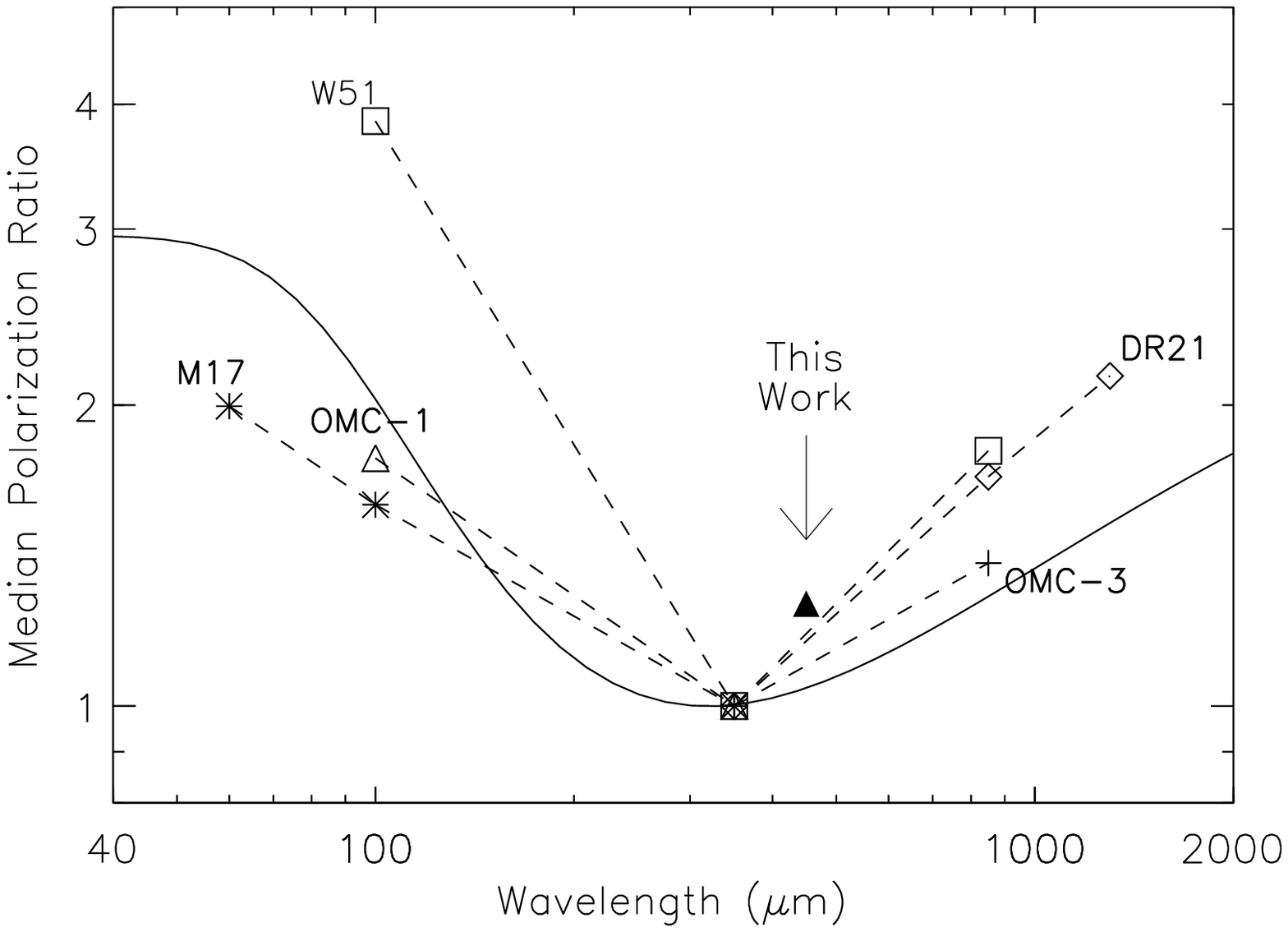}
\caption{Far-infrared and submillimeter polarization spectrum,
  normalized at $350\,\micron$. The 450/350 $\micron$ OMC-1 comparison
  from this work is shown as a solid triangle.  The 850/350 $\micron$
  comparison in W51 (open squares) is calculated from $850\,\micron$
  data in \citet{w51scuba} and $350\,\micron$ data in
  \citet{hertzarchive}.  All other data are from \citet{mythesis}.
  The solid curve is a 2-component dust model (see text).}
\label{fig-pspec}
\end{figure}

Examining SHARP data at only points outside the KL and KHW intensity
peaks (beyond a $20\arcsec$ radius) we find a median in the
distribution of $P(450)/P(350) = 1.3 \pm 0.3$, where the uncertainty
represents the standard deviation of those points (Fig.\
\ref{fig-hist}).  This data point is consistent with a simple
interpolation of data from other Galactic clouds at wavelengths of 350
and 850 $\micron$ (Fig.\ \ref{fig-pspec}).  (Note that Fig.\
\ref{fig-pspec} also includes data at $850\,\micron$ from the Galactic
cloud W51 \citep{w51scuba}, not available at the time of the
\citet{mythesis} compilation; the comparison between the
$850\,\micron$ SCUBA data and the $350\,\micron$ Hertz data
\citep{hertzarchive} is performed as outlined in \citet{mythesis}.)
From the results of Figure \ref{fig-pspec}, and assuming that the
polarization spectrum in OMC-1 is similar to that of other clouds
(i.e., it continues to rise at wavelengths greater than
$450\,\micron$), we estimate that the minimum in the spectrum cannot
occur at wavelengths much less than $350\,\micron$.

\subsection{Polarization Ratio in Different Environments}

\begin{figure}[t]
\plotone{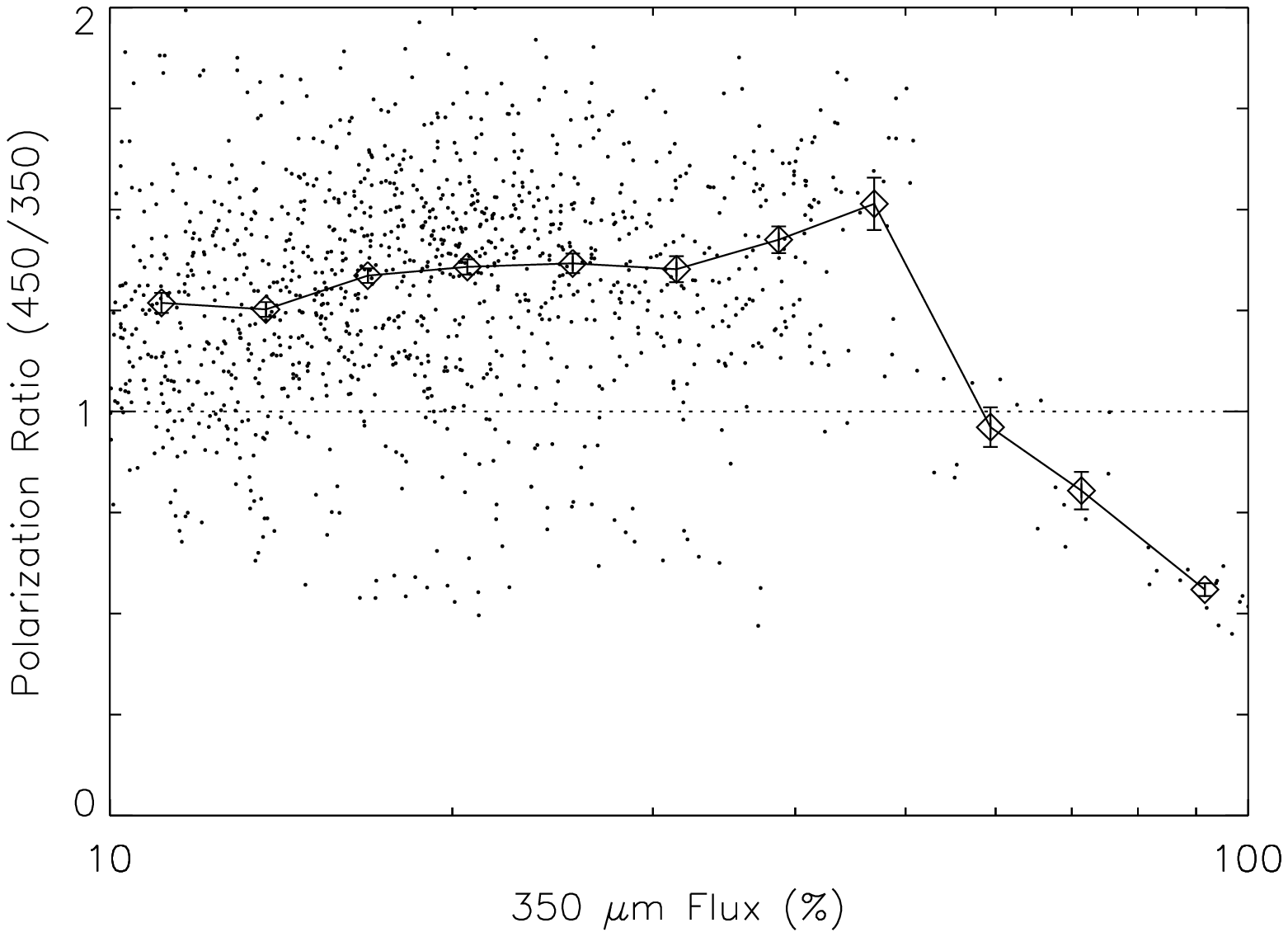}
\caption{Polarization ratio (450 / 350 $\micron$) vs.\ the total
  $350\,\micron$ flux (percentage of peak flux).  Dots indicate
  individual measurements within the maps of Figure
  \ref{fig-map1}\emph{c} while diamonds represent an average of the
  plotted quantity over logarithmic bins of $350\,\micron$ flux. Error
  bars on the polarization ratio represent the standard deviation of
  the mean within each bin.}
\label{fig-pvsf}
\end{figure}

Given the low polarization ratio towards KL compared with the rest of
the cloud, it is natural to ask if this may be due to some unique
physical condition as compared with the remainder of the cloud. At the
very least, it is obvious that KL is the region with the highest flux
density in the cloud.  To examine this trend, we plot the 450/350
polarization ratio as a function of $350\,\micron$ flux
$F_{350}$. Figure \ref{fig-pvsf} plots the polarization data as
individual points (dots) and combined into equal-sized
logarithmic flux bins (diamonds).  At low fluxes ($F_{350} \lesssim
50\%$ of the peak flux) the polarization ratio clusters about
$P(450)/P(350) \approx 1.2$ -- 1.5 with a large scatter (this
distribution is best shown in the histogram of Figure \ref{fig-hist}).
At larger fluxes the binned data show a sharp drop in the polarization
ratio.  The flux level at which this drop occurs is indicated by the
thick contours in Figure \ref{fig-map1}\emph{c}.  With few exceptions
all data above this flux level lie in KL\@. The binned data in
Figure \ref{fig-pvsf} drop below $P(450)/P(350) = 1$ for $F(350)
\gtrsim 60$\%.  At this flux level all points are within $\sim
20\arcsec$ of KL.

Polarized emission from dust grains all with the same temperature
results in a nearly constant polarization spectrum in the wavelength
range 50 -- 2000 $\micron$ \citep{pspec}.  A spectrum such
as that in Figure \ref{fig-pspec}, or points in Figure \ref{fig-pvsf}
with $P(450)/P(350) \neq 1$, requires the existence of at least two
dust-emission components \citep{pspec,mythesis}.  Such a polarization
spectrum is modeled in Figure \ref{fig-pspec}.  This model has a
minimum at $330\,\micron$ and uses optically thin dust with
temperatures $T =$ 20 and 50 K, spectral indices $\beta =$ 2 and 1,
respectively,\footnote{An inverse $T$-$\beta$ relation is observed to
  be typical for dust in molecular clouds (e.g.,
  \citealt{pronaos-m42,pronaos-compile}).} and peak fluxes in the
ratio $F_\mathrm{cold} / F_\mathrm{hot} = 0.5$.  The cold component is
unpolarized while the warm component is polarized.  While a
two-component dust model for molecular clouds is physically
unrealistic, it is sufficient to illustrate how a multi-component
model can explain the empirical results.

Even with only 2 components, small adjustments to the parameters allow
for better agreement between the model and the specific data points of
Figure \ref{fig-pspec}.  A spectral energy distribution (SED)
dominated by cold unpolarized dust ($F_\mathrm{cold} / F_\mathrm{hot}
\gg 1$) can reproduce the ratio observed in the OMC-1 cloud envelope,
$P(450)/P(350) \approx 1.3$. Such an SED will appear nearly-isothermal
if the measurement uncertainties are large and/or it is not
sufficiently sampled in wavelength space.  This is consistent with the
analysis of \citet{mythesis} where no significant evidence is seen for
multiple dust temperatures outside of KL or the M42 \ion{H}{2} region,
but where the polarization spectrum is still observed to change with
wavelength.

The KL region is quite massive and contains high column densities of
both warm ($T \gtrsim 40$\,K) and cold ($T \lesssim 25$\,K) dust. As a
result, emission from cold and warm dust contribute nearly equally to
the total flux observed at 350 and 450 $\micron$.  The coldest
($T\approx 20$\,K) and warmest ($T\approx45$\,K) regions of
\mbox{OMC-1} are regions towards KL and KHW \citep{mythesis}.
As one moves towards KL from within $30\arcsec$ of the peak the
temperature of the cold component decreases by $\sim 2$\, K, and the
warm component increases by $\sim 3$\,K\@.  These modest temperature
changes are enough to shift the polarization minimum by as much as 70
-- 130 $\micron$ (depending on the relative spectral indices of the
components); moving to longer wavelengths towards the intensity peak.  

Therefore, we expect the polarization minimum to shift to longer
wavelengths at the larger flux values towards the KL peak.  If the
observed wavelengths are near the minimum then we expect the 450/350
ratio to decrease from a value greater than unity when the minimum is
below $400\,\micron$, to a value approximately equal to one when the
minimum is bracketed by the two observed wavelengths, and to continue
to decrease below one as the minimum moves to wavelengths greater than
$400\,\micron$.  This is the trend observed in Figure \ref{fig-pvsf}
for points with $F_{350} \gtrsim 50\%$.

\subsection{Summary and Future Work}

By comparing the 450/350 $\micron$ polarization ratio in OMC-1 with
that in other bright Galactic clouds (Fig.\ \ref{fig-pspec}) we
estimate that the minimum in the polarization spectrum is not much
less than $350\,\micron$.  However, given the variation in 60 -- 1300
$\micron$ polarization spectrum from cloud-to-cloud, we do not draw
strong conclusions about the polarization minimum in any cloud other
than OMC-1.  We are continuing our campaign to observe and analyze
other bright clouds in a manner similar to that described here.

The existence of a minimum in the far-infrared/submillimeter
polarization spectrum, as opposed to a simple rise or fall from one
end of the spectrum to the other, implies either (a) the existence of
more than 2 dust temperature components, and/or (b) a change in the
dust emissivity index as well as its temperature.  Multi-wavelength
observations in the far-infrared (50 -- 200 $\micron$;
\citealt{archive,hale,spie2007}) or submillimeter ($> 450\,\micron$;
\citealt*{archivescuba,scuba2}; Matthews et al., in preparation) at
wavelengths on only one side of this minimum are insufficient for
studying this behavior.  SHARP (at 350 and 450 $\micron$) therefore
occupies a unique niche for studies of the polarization spectrum by
providing a link between these two wavelength extremes.


\acknowledgments

We are grateful for the help of the Caltech Submillimeter Observatory
(CSO) staff in installing and observing with SHARP and SHARC-II\@.
Also thanks to Jackie Davidson, and Emeric Le\,Floc'h for observing
assistance.  SHARP has been supported by NSF grants AST 02-41356, AST
05-05230, and AST 05-05124.  The CSO is supported by the NSF through
grant AST 05-40882.

{\it Facilities:} \facility{CSO (SHARC2)}.

\bibliography{hildebrand,me,bolometer,other,polar,tmp,dust}
\bibliographystyle{apj}


\clearpage
\begin{figure}
\centering
\includegraphics[width=5.5in]{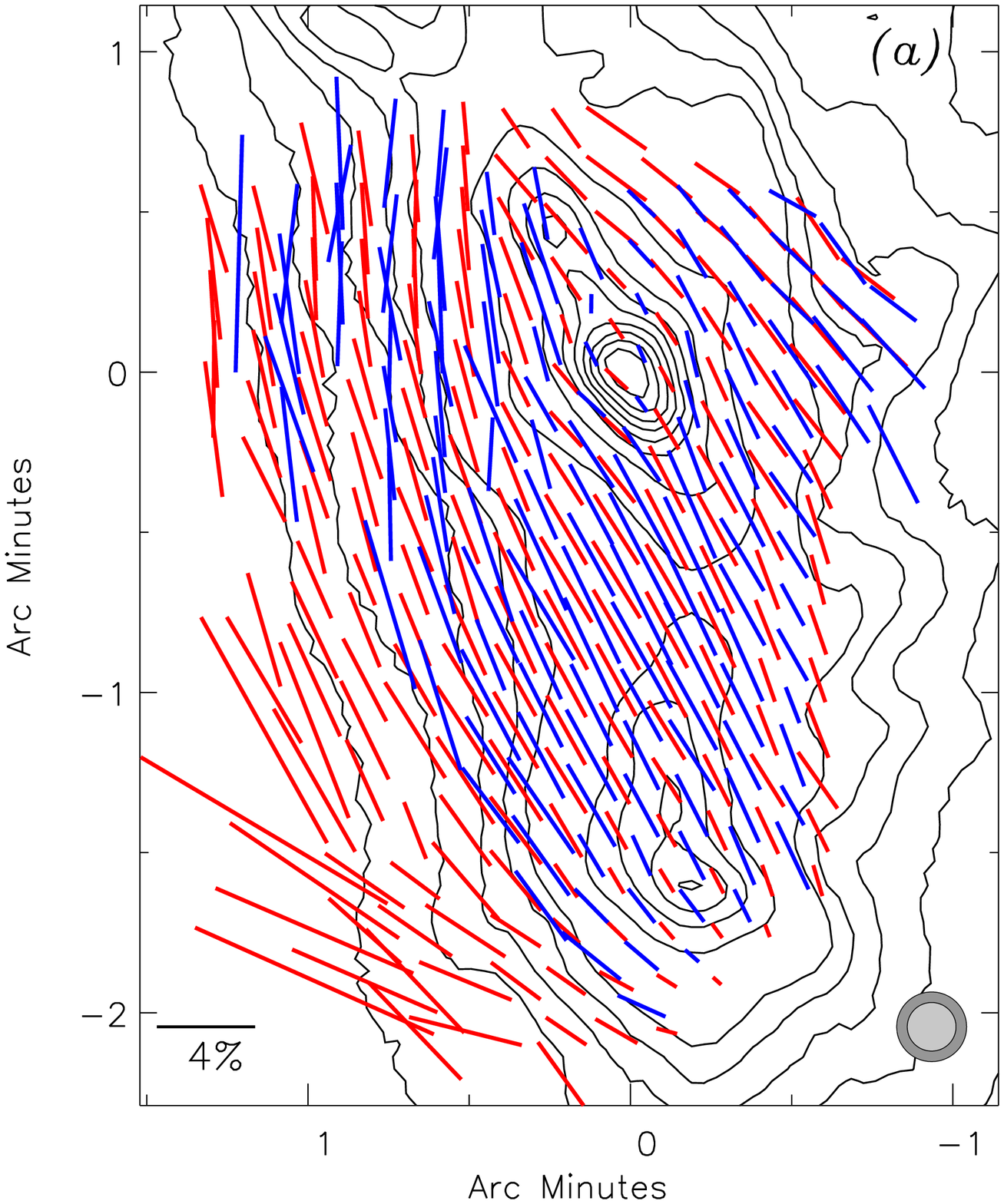}
\vskip 0.2in
\emph{\large Figure 1a}
\end{figure}

\clearpage
\begin{figure}
\centering
\includegraphics[width=5.5in]{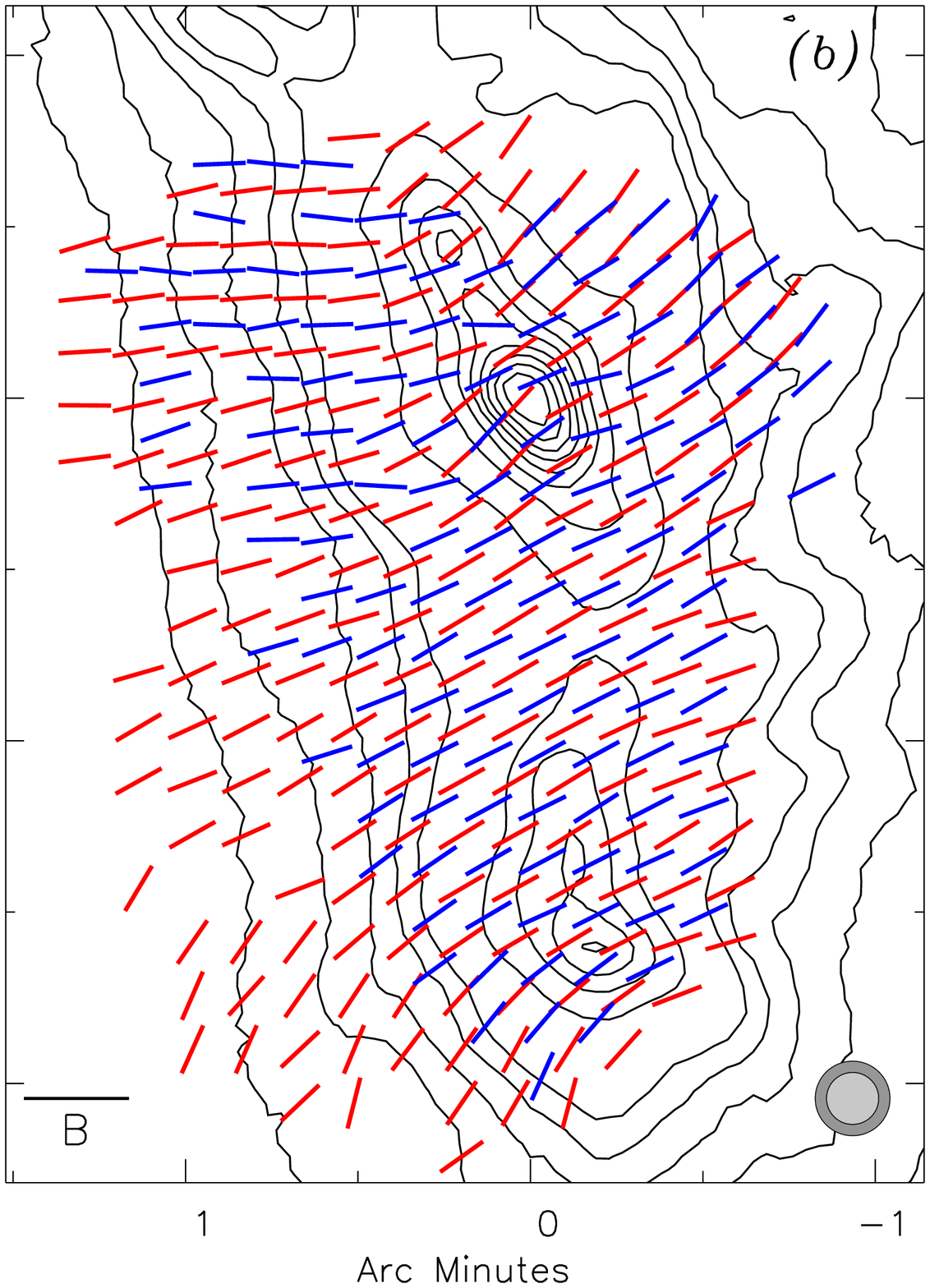}
\vskip 0.2in
\emph{\large Figure 1b}
\end{figure}

\clearpage
\begin{figure}
\centering
\includegraphics[width=5.5in]{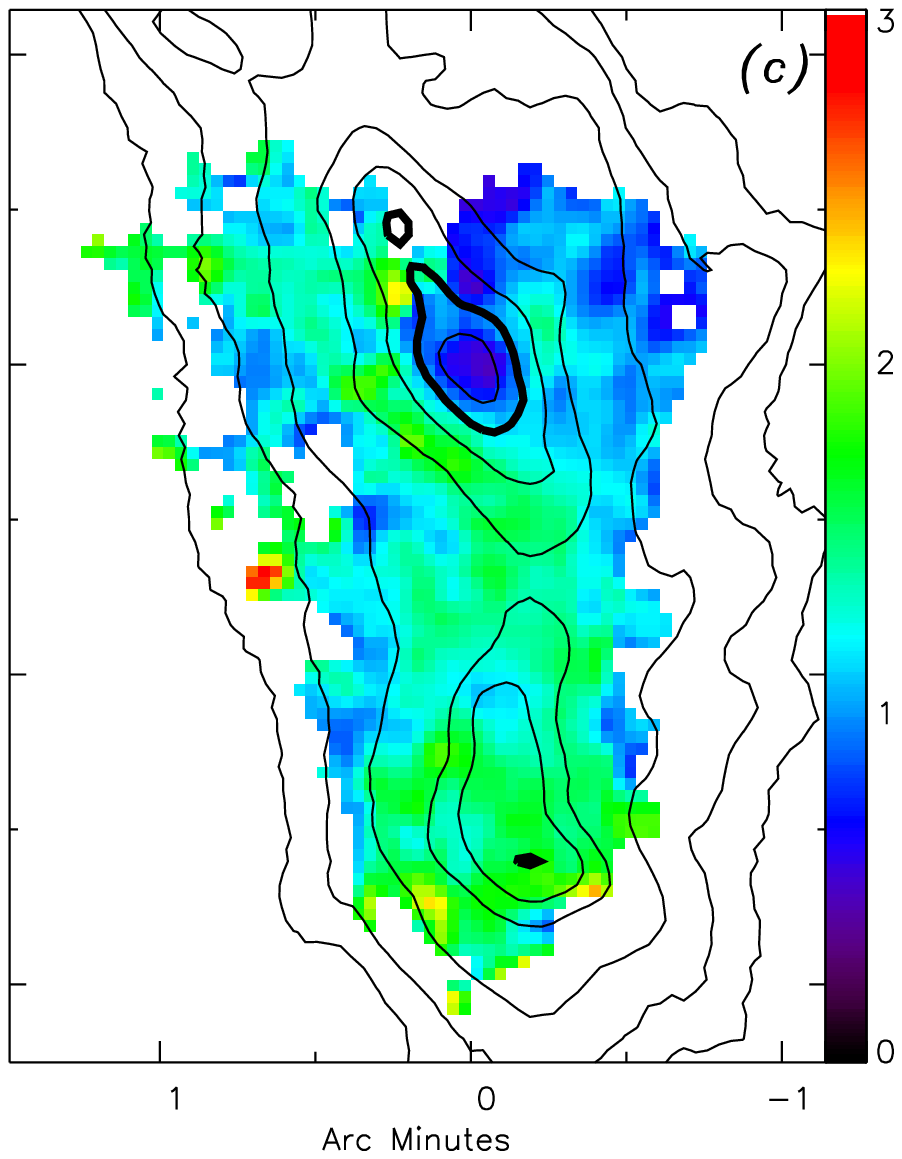}
\vskip 0.2in
\emph{\large Figure 1c}
\end{figure}

\setcounter{figure}{0}
\begin{figure*} \centering
\caption{Polarimetric and photometric maps of OMC-1.  Effective
  beamsizes (FWHM) for the photometric ($9\arcsec$) and polarimetric
  ($13\arcsec$) observations are shown as gray circles in \emph{(a)}
  and \emph{(b)}. Coordinate offsets are measured with respect to
  Ori\,IRC2 at $5\hour 35\minute 14.5\second$, $-5\arcdeg 22\arcmin
  31\arcsec$ (J2000). KL is the northernmost flux peak coincident with
  the coordinate origin and KHW/Orion-south is the peak $\sim
  1\farcm5$ to the south.  Only polarization data satisfying $P >
  3\sigma_p$ are included. \emph{a}) 350 (red) and 450 (blue)
  $\micron$ polarization vectors superposed on 350 $\micron$ flux
  contours. Contours are drawn at 2, 4, 6, 8, 10, 20, ..., and 90\% of
  the peak ($\approx 780$ Jy per $9\arcsec$ beam). \emph{b}) Inferred
  magnetic field vectors at 350 and 450 $\micron$ drawn with a
  constant length (i.e., not proportional to the polarization
  amplitude); contours as in (\emph{a}). \emph{c})~The color scale
  shows the polarization ratio between the two wavelengths,
  $P(450)/P(350)$. Contours at $350\,\micron$ are drawn at 4, 6, 10,
  20, 30, 50, and 80\% of the peak flux; the 50\% contour is drawn
  thicker (see Fig.\ \ref{fig-pvsf}).}
\label{fig-map1}
\end{figure*}

\end{document}